\documentclass[journal,10pt]{IEEEtran}

\usepackage{multirow}
\usepackage{amsmath}
\usepackage{graphicx}
\usepackage{optidef}
\usepackage{amsfonts}
\usepackage{lipsum}

\usepackage{mathtools}
\usepackage{cuted}
\usepackage{soul}
\usepackage{xcolor}

\begin{document}
\title{NB-IoT via LEO satellites: An efficient resource allocation strategy for uplink data transmission}

\author{Oltjon~Kodheli, Nicola~Maturo, Symeon~Chatzinotas, Stefano Andrenacci, Frank Zimmer 
	
	\thanks{This work was supported by the
		Luxembourg National Research Fund (FNR)
		under Industrial Fellowship Scheme, SATIOT project, Grant FNR12526592.}
	
	\thanks{O. Kodheli, N. Maturo and S. Chatzinotas are with Snt - Interdisciplinary Centre for Security, Reliability and Trust, University of Luxembourg, 29 Avenue J.F. Kennedy,Luxembourg City L-1855, Luxembourg (e-mail: \{oltjon.kodheli, nicola.maturo, symeon.chatzinotas\}@uni.lu)}

    \thanks{S. Andrenacci and F. Zimmer are with SES S.A. (e-mail: \{stefano.andrenacci, frank.zimmer\}@ses.com)}}


\mark{This work has been submitted to the IEEE IoT Journal for possible publication. Copyright may be transferred without notice, after which this version may no longer be accessible.}
{}


\maketitle

\begin{abstract}
In this paper, we focus on the use of Low-Eart Orbit (LEO) satellites providing the Narrowband Internet of Things (NB-IoT) connectivity to the on-ground user equipments (UEs). Conventional resource allocation algorithms for the NB-IoT systems are particularly designed for terrestrial infrastructures, where devices are under the coverage of a specific base station and the whole system varies very slowly in time. The existing methods in the literature cannot be applied over LEO satellite-based NB-IoT systems for several reasons. First, with the movement of the LEO satellite, the corresponding channel parameters for each user will quickly change over time. Delaying the scheduling of a certain user would result in a resource allocation based on outdated parameters. Second, the differential Doppler shift, which is a typical impairment in communications over LEO, directly depends on the relative distance among users. Scheduling at the same radio frame users that overcome a certain distance would violate the differential Doppler limit supported by the NB-IoT standard. Third, the propagation delay over a LEO satellite channel is around 4-16 times higher compared to a terrestrial system, imposing the need for message exchange minimization between the users and the base station. In this work, we propose a novel uplink resource allocation strategy that jointly incorporates the new design considerations previously mentioned together with the distinct channel conditions, satellite coverage times and data demands of various users on Earth. The novel methodology proposed in this paper can act as a framework for future works in the field.

\end{abstract}


\IEEEpeerreviewmaketitle

\section{Introduction}

The Narrowband Internet of Things (NB-IoT) standard is expected to play a crucial role in the fifth generation mobile communication (5G) network. Since its inception, the NB-IoT was envisioned to address the massive machine type communication (mMTC) traffic, broadening the scope of 5G applications towards smart cities, smart homes, connected cars, smart meters, etc. The terrestrial-only networks fail to guarantee global coverage of services due to their inability to cover certain areas. In such areas, where there is no possibility or economic advantage in building a terrestrial infrastructure, the satellite communication systems (SatComs) are inevitable. As a matter of fact, after an initial study phase \cite{ref10}, it is now approved by the 3rd Generation Partnership Project (3GPP) that satellites will be a new key feature of the 5G, and a work item (WI) has already started for Release 16 and 17 \cite{ref1}. NB-IoT via satellite systems are also foreseen to be included in the study, potentially targeting Release 17 \cite{ref11}. The study will focus on the adaptations of the NB-IoT protocol so as to counteract the satellite channel impairments, which are very different compared to the terrestrial ones. 

\subsection{Motivation and Related Work}

\subsubsection{NB-IoT via LEO satellite}

Several works in the literature have studied satellite-based NB-IoT systems. The role that the satellite can play in the 5G IoT communications has been analyzed in \cite{ref12, ref13}. Furthermore, the authors in \cite{ref14} assess the impact of large delays and Doppler effects in two scenarios of interest for future 5G NTN systems, enhanced Mobile BroadBand (eMBB) services and NB-IoT. In particular, for the NB-IoT scenario, they investigate the challenges that the increased delay and differential Doppler shift in a LEO satellite channel impose in the physical (PHY) and medium access control (MAC) layer procedures. To decrease the amount of differential Doppler down to a limit supported by the NB-IoT devices, in our previous work \cite{ref9}, we propose a radio resource allocation which mainly relies on a group-based data transmission strategy. Furthermore, in \cite{ref18}, we validate our technique through numerical simulations. However, the proposed resource allocation strategy is limited since it does not take into account other crucial parameters when assigning radio resources to various users, such as different data volumes, distinct channel conditions or any other technical peculiarity related to the satellite channel (e.g. system dynamicity or increased delay). Besides, when designing a resource allocation strategy for NB-IoT systems, it is highly important to optimize the available radio resources, which in the NB-IoT case are very limited (only 180 kHz of bandwidth available). Alternative works exist, targeting strategies on how to optimally share the radio resources among different satellites in a heterogeneous satellite network \cite{ref32, ref33}, but without targeting a specific use case or application. To the best of our knowledge, no other work in the literature has analyzed and proposed a radio resource allocation method explicitly applicable to NB-IoT systems over a LEO satellite. 

\subsubsection{Radio resource allocation for terrestrial NB-IoT}

Prior works on radio resource allocation for NB-IoT terrestrial networks have addressed different issues. In \cite{ref19}, an algorithm that predicts the processing delay and pre-assigns radio resources for uplink data transmission has been proposed. Through this strategy, the number of random access attempts can be diminished, consequently reducing the battery consumption of the NB-IoT devices. In \cite{ref20}, inter-cell interference aware radio resource allocation to devices in uplink and downlink NB-IoT connectivity has been investigated, where the aim is to maximize the achievable data rate of each cell. In \cite{ref21}, an uplink resource allocation algorithm is proposed that dynamically selects the modulation and coding scheme (MCS) and channel repetition number based on estimated real-time channel state information (CSI). It has been shown that the proposed mechanism reduces the radio resource consumption and is particularly efficient for users with good channel conditions and large packet sizes. Another uplink resource allocation strategy with the objective of minimizing the consumed radio resources, has been presented in \cite{ref22}. The proposed algorithm is able to satisfy different requirements of various users in terms of the volume of data to be transmitted. In contrast to the above mentioned works, which particularly target the NB-IoT uplink, in \cite{ref23}, a scheduling strategy has been considered accounting for both, the uplink and downlink NB-IoT channels. An analytical model of the channel scheduling problem has been developed and used to analyze the performance trade-offs in terms of the expected latency and battery lifetime of the devices in various channel conditions. 

The above-mentioned algorithms are particularly designed for a terrestrial NB-IoT network, where devices are under the coverage of a specific base station, and are grouped in three different coverage classes based on their distance to the base station. The proposed strategies target the optimization of one ore more resources, such as minimizing the energy consumption, minimizing the radio resource utilization or maximizing the data rate, while accounting for various channel conditions and quality of service requirements of users placed in different coverage classes. Developing a resource allocation algorithm for a LEO-satellite based NB-IoT system would require a distinct approach taking into account other design considerations. This is due to the following reasons.
	
\paragraph{Extremely dynamic system} With the movement of the LEO satellite, the corresponding channel parameters for each user will change over time. One of these important parameters is the distance to the satellite, which would cause the users to jump from one coverage class to another with the satellite movement. In addition to this, the coverage provided by the LEO satellite to the on-ground users would be limited in time and diverse for each of them. Consequently, in such a dynamic and heterogeneous system, there is a need to take proactive decisions regarding the users to schedule in the same frame and how to efficiently distribute the radio resources among them. Delaying the scheduling of a certain user would result in a  resource allocation based on outdated parameters, thus impacting its efficiency in optimizing the available resources. On the contrary, in a terrestrial NB-IoT network, such time-restriction does not exist, because the system is static or it varies very slowly in time.
	
\paragraph{Low complexity requirement} The responsible entity for assigning resources, both in downlink and uplink transmission, is the base station. Since a LEO-satellite based NB-IoT system would be highly dynamic, requiring frequent and proactive decisions regarding the resource allocation, it brings the need for practical algorithms. Besides, this would also be fundamental for scenarios with flying base stations where the processing would be done on-board the satellite. Please note that in a terrestrial NB-IoT network such constraint is not present because the system is quasi-static and there is no power limitation at the base station side. 
	
\paragraph{Increased round trip delay (RTD)} The RTD represents the propagation time that the signal takes to be transmitted from the user to the base station and vice versa. In a terrestrial NB-IoT network, accounting for the largest cell size, the range of RTD is lower than 1 ms. In contrast, the RTD over a LEO satellite channel would be significantly higher, increasing the delay in the overall NB-IoT system. Clearly, the RTD would depend on the specific LEO orbit, ranging from 4-16 ms. As a consequence, it would be beneficial to minimize the message exchange between the users and the base station to counterbalance the increased RTD.  
	
\paragraph{High differential Doppler among users} As it has been shown in \cite{ref14}, the differential Doppler shift overcomes the supported limit by the NB-IoT standard even at low carrier frequencies, or higher LEO altitudes. This is a very essential problem in the PHY layer of communication which needs to be solved, because it results in high inter-carrier interference and significantly impacts the NB-IoT performance \cite{ref18}. As we have analyzed in \cite{ref9}, it is possible to tackle the differential Doppler problem through a resource allocation strategy. This can be done by scheduling at the same radio frame only users that do not violate the differential Doppler limit. The advantage of this approach is that it is highly practical and it does not increase the complexity at the user side, but only at the base station side. Other works in the literature propose a completely new air interface, able to tackle the LEO satellite channel impairments, including the Doppler effects among others \cite{ref15, ref17}. However, following this approach requires totally new chipsets for the NB-IoT, which goes against the concept of standardization.

\subsection{Paper Contributions and structure}

In this paper, we design a novel uplink resource allocation strategy, which jointly incorporates the technical peculiarities of the LEO satellite channel with the various channel conditions and different user requirements in terms of data to be transmitted. To the best of our knowledge, this is an untreated problem in the literature and the methodology proposed in this paper can act as a framework for future works in this area. More specifically, the main contributions of this paper can be summarized as follows:
\begin{itemize}
	
    \item We mathematically model the user scheduling strategy as a 0-1 Two-dimensional Knapsack Problem. Solving such a problem, means selecting the set of users to transmit their uplink data in the available radio resources which maximizes the sum profit of the selected users.		
		
	\item We formulate a profit function which assigns different profits to users depending on their data packet sizes, channel conditions and satellite visibility time. The latter one is a unique and crucial feature, while the data volumes and channel conditions have already been contemplated in a terrestrial scheduling.
	
	\item By fixing the transmission mode only to single-tone, we convert the problem into a simpler 0-1 Multiple Knapsack Problem (MKP). Exploiting the unique features of our system, we propose an approximate algorithm with a near-optimal performance and low complexity.
	
	\item We study the performance trade-offs of our LEO satellite-based NB-IoT system through numerical simulations by changing the weights of the components which contribute in the profit function of the 0-1 MKP problem. This allows us to find optimal operation points for three different Key Performance Indicators (KPIs) and rank various algorithms.
\end{itemize}

The remainder of the paper is structured as follows. In Section \ref{sec2} we make a brief overview of the responsible entities for the NB-IoT radio resource management. In Section \ref{sec3} we describe the NB-IoT via LEO satellite system model and assumptions. Section \ref{sec4} and \ref{sec5} are dedicated to the mathematical formulation of our resource allocation strategy and the corresponding solutions. In Section \ref{sec6} we evaluate the performance trade-offs of our system through performing numerical simulations and in Section \ref{sec7} we discuss the conclusions and the future work. 

\section{The management of radio resources in NB-IoT} \label{sec2}

This section aims to describe how the radio resource management is handled in an NB-IoT system and identify the responsible entities for various functionalities in the management process. This is done to justify and better explain the assumptions in our system modeling, avoiding the technical details which are not important for the purpose of this work. 

\subsection{Physical channels and signals}

\begin{figure}[!t]
	\centering
	\includegraphics[width = 0.5\textwidth]{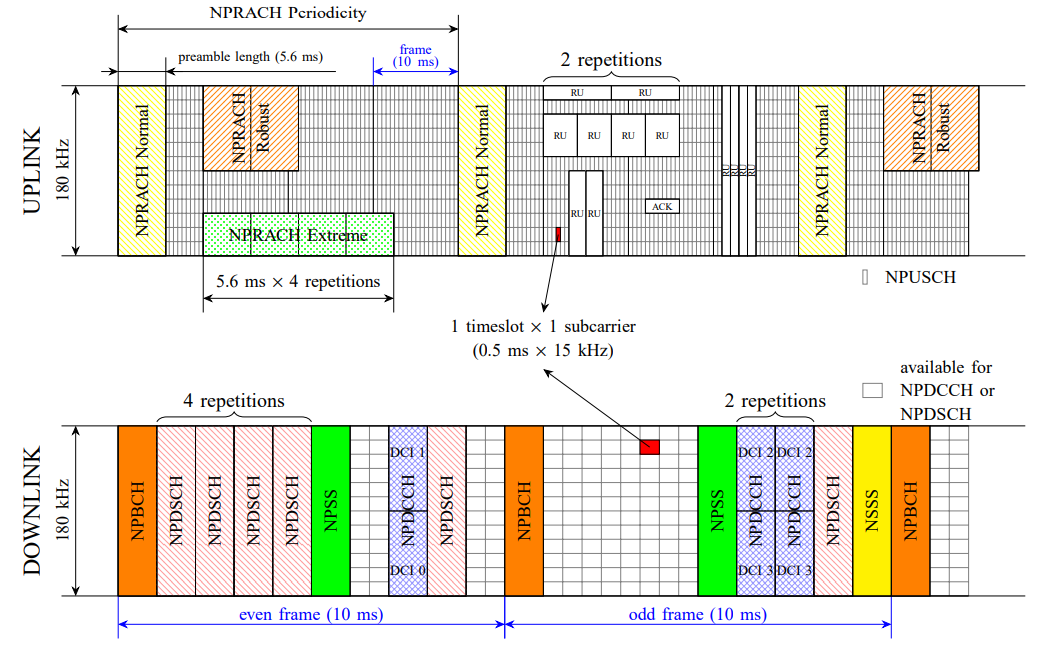}
	\caption{An example of physical channels in the DL and UL frame \cite{ref24}.}
	\label{fig1}
\end{figure}

There are various PHY channels and signals that an NB-IoT UE uses, and each of them is intended to play a specific role in the whole NB-IoT system. Also, the resources occupied by these channels and signals can be different. Fig. \ref{fig1} shows an illustrative example of the time-frequency multiplexing of various channels in the downlink and uplink radio frames, taken from \cite{ref24}. Hereafter, we will explain their functionalities considering the three main operational phases of a certain UE: synchronization, access and data phase.

\subsubsection{Synchronization Phase}

The NB-IoT employs a strategy where UEs periodically go into sleeping mode for the purpose of extending their battery lifetime. When a UE has data to transmit after a period of inactivity, it needs to firstly synchronize in time and frequency with the serving base station (BS). For this purpose, it utilizes the narrowband physical primary synchronization signal (NPSS) and narrowband secondary synchronization signal (NSSS). Basically, after achieving the time-frequency synchronization with the carrier frequency of the serving BS, the UE is able to decode the narrowband physical broadcast channel (NPBCH). The NPBCH is transmitted at the beginning of each radio frame and contains the useful information needed by the UE in order to proceed with the next phase, the access phase.   

\subsubsection{Access Phase}

To access the NB-IoT network, the UEs has to perform the random access (RA) procedure, which relies on a 4-step message exchange detailed below.

\begin{figure}[!t]
	\centering
	\includegraphics[width = 0.35\textwidth]{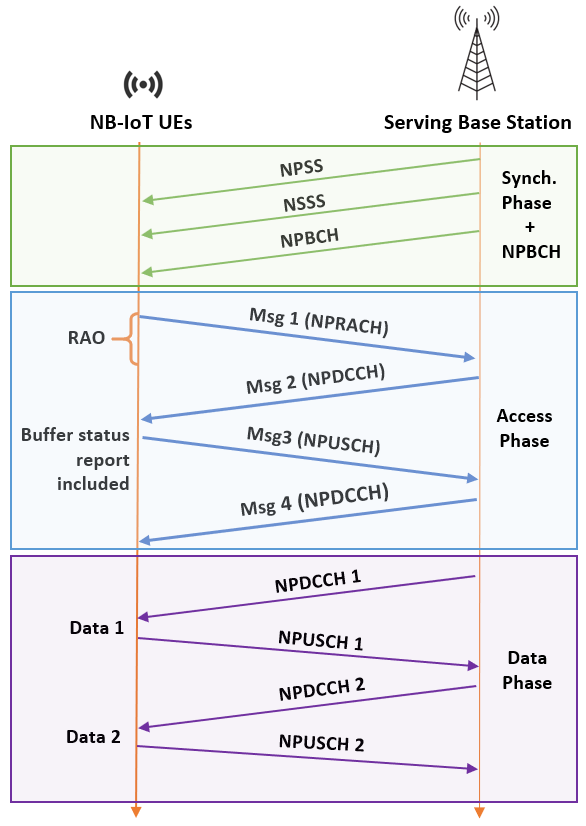}
	\caption{An example of message exchange between BS and UE.}
	\label{fig2}
\end{figure}

\paragraph{Message (msg) 1} The UE sends a random access preamble using the narrowband physical random access channel (NPRACH). There exist three NPRACH configurations, corresponding to three coverage enhancement (CE) levels defined in the standard \cite{ref24}, which mainly depend on the distance of the UEs from the BS. In Fig. \ref{fig2} various NPRACH configurations are illustrated. The specific details for each NPRACH configuration, meaning the time-frequency resources and periodicity, is reported in the NPBCH. This is why the UE cannot initiate the RA procedure without decoding the NPBCH channel.
\paragraph{Msg 2} When the UEs send the preamble though Msg 1, they randomly select from a set of given frequencies inside the NPRACH channel. If two or more UEs select the same frequency for preamble transmission, a collision occurs. The un-collided preambles will be detected by the BS, which then sends a RA response (RAR) message, known as Msg 2. 
\paragraph{Msg 3} Through the msg 3, the UE notifies the BS that it wishes to transmit and reports its buffer status. Through this information and the power estimate of the received signal, the BS can take proper decisions for the radio resource allocation of various users.
\paragraph{Msg 4} In the final step, the BS assigns a permanent ID to the UE, which now will be in a connected mode and is ready to transmit its uplnik data. 

\subsubsection{Data Phase}

Once the user has a successful access phase, it can transmit its uplink data using the narrowband physical uplink shared channel (NPUSCH), or receive the downlink data utilizing the narrowband physical downlink shared channel (NPDSCH). Please note that in the downlink, the smallest unit that can be assigned to a user is the physical resource block (PRB), corresponding to 12 subcarriers (12 x 15 kHz = 180 kHz) in the frequency domain and 2 slots in the time domain. On the other hand, in the uplink case, the smallest schedulable unit is the resource unit (RU). There are various representations of the RU, occupying different time-frequency resources. In terms of frequency, the RU can occupy 1, 3, 6 or 12 subcarriers corresponding to a time duration of 8 ms, 4 ms, 2 ms and 1 ms respectively. When the RU utilizes only 1 subcarrier (15 kHz), it is known as a single-tone transmission, whereas the other configurations as multi-tone. Since the downlink channels occupy the whole available bandwidth, in the downlink frames there exist only a time-multiplexing of channels. The same does not hold for the uplink due to the various RU configurations, allowing also for frequency multiplexing of different users in the same frame.
\begin{table}[t]%
	\caption{Transport block size (TBS) table for NPUSCH \cite{ref5}.}\label{tab1}
	\centering
	\setlength{\tabcolsep}{4pt}
	\begin{tabular*}{226pt}{|c||c|c|c|c|c|c|c|c|c|}
		\hline
		\multicolumn{1}{|c||}{\textbf{MCS level}} &
		\multicolumn{8}{c|}{\textbf{Number of RUs}} \\
		\cline{2-9}
		& 1 & 2 & 3 & 4 & 5 & 6 & 8 & 10 \\
		\hline
		\hline
		0 & 16 & 32 & 56 & 88 & 120 & 152 & 208 &256\\
		\hline
		1 & 24 & 56 & 88 & 144 & 176 & 208 & 256 & 344 \\
		\hline
		2 & 32 & 72 & 144 & 176 & 208  & 256 & 328 & 424 \\
		\hline
		3 & 40 & 104 & 176 & 208 & 256  & 328 & 440 & 568 \\
		\hline
		4 & 56 & 120 & 208 & 256 & 328  & 408 & 552 & 680\\
		\hline
		5 & 72 & 144 & 224 & 328 & 424  & 504 & 680 & 872\\
		\hline
		6 & 88 & 176 & 256 & 392 & 504  & 600 & 808 & 1000 \\
		\hline
		7 & 104 & 224 & 328 & 472 & 584  & 712 & 1000 & 1224\\
		\hline
		8 & 120 & 256 & 392 & 536 & 680  & 808 & 1096 & 1384\\
		\hline
		9 & 136 & 296 & 456 & 616 & 776  & 936 & 1256& 1544\\
		\hline
		10 & 144 & 328 & 504 & 680 & 872  & 1000 &1384 &1736\\
		\hline
		11 & 176 & 376 & 584 & 776 & 1000  & 1192 &1608 &2024 \\
		\hline
		12 & 208 & 440 & 680 & 1000 &  1128 & 1352 &1800 &2280 \\
		\hline
		13 & 224 & 488 & 744 & 1032 &  1256 & 1544 & 2024& 2536\\
		\hline
	\end{tabular*}
\end{table}

\subsection{Physical Resource Orchestration}

The responsible entity for orchestrating the available radio resources among various users is the BS. The users are informed with the corresponding parameters regarding the assigned resources by decoding the information encapsulated in the narrowband physical downlink control channel (NPDCCH). For the NPUSCH transmission, the most important ones worth mentioning here are the number of subcarriers to be utilized, the number of RUs, the modulation and coding scheme (MCS), and the specific subframe to start the transmission. Obtaining this information, the user will know the time-frequency resources for its NPUSCH transmission, the time-duration of the NPUSCH (depending on the number of RUs) and the number of bits to encode in the NPUSCH depending on the assigned MCS level and number of RUs. Table \ref{tab1} shows the possible transmission block sizes (TBS) to be encoded in the NPUSCH by the UE. There are various MCS levels that the BS may assign to a certain user based on its specific channel conditions. In addition, depending on the buffer status report, more than one RU can be assigned for NPUSCH transmission. Clearly, the higher the MCS level (better channel conditions), the more bits can be encoded in the same NPUSCH duration, and consequently the higher would be the data rate. Last but not least, it is possible that the buffer of a certain user is emptied through several NPUSCH transmissions. Obviously, this will require several NPDCCH as well to provide the necessary information regarding the resources to be used. The resource allocation strategy is open for design by the operator, but it has to follow the above-mentioned standard specifications. For the NPDSCH transmission the same reasoning applies. The only difference is that the BS does not need to specify the number of subcarriers to be utilized since in the downlink the whole NB-IoT carrier is assigned only to one channel. As a result, the NPDSCH for various users is transmitted in different subframes (time-division multiplexing). Fig. \ref{fig2} shows an example of the message exchange between the BS and the UE until its uplink data is offloaded.

\section{NB-IoT via LEO satellite: System model \& assumptions} \label{sec3}

\begin{figure}[!t]
 	\centering
 	\includegraphics[width = 0.5\textwidth]{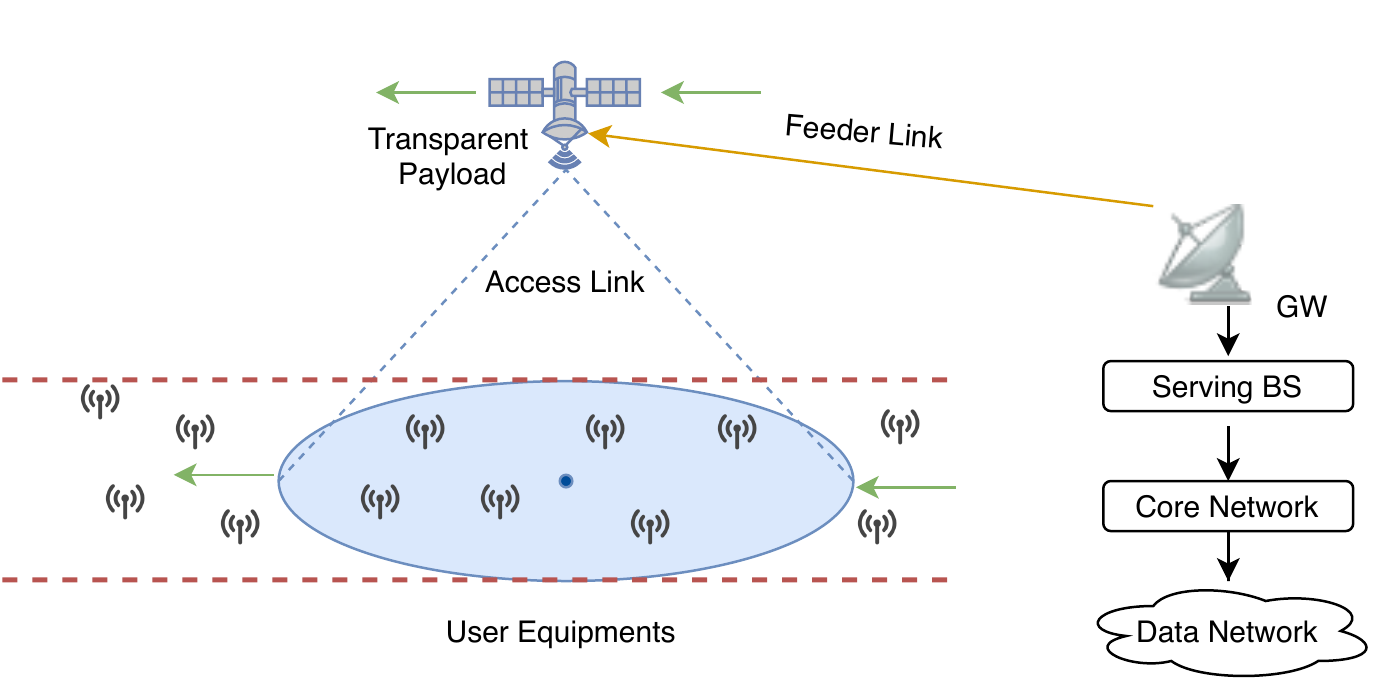}
 	\caption{NB-IoT via LEO satellite system.}
 	\label{fig3}
 \end{figure}

Regarding the system model, we consider a LEO satellite with a transparent payload, providing NB-IoT coverage to a certain number of on-ground UEs. The satellite acts as a relay providing the link between the UEs and the on-ground serving BS through the Gateway (GW). The air interface in the access and the feeder link is the same as the terrestrial one, based on the 4G Uu air interface or 5G New Radio (NR) one. This is because the NB-IoT protocol was initially standardized to be compatible with the 4G network, and later it was agreed to address the IoT traffic of the 5G as well. In either case, the technical specifications for the NB-IoT would be the same, only with different terminology for the network components. The BS is then connected to the 4G/5G Core and Data Network. Last but not least, we consider an Earth-moving cell, hence creating a dynamic system with users continuously entering/leaving the coverage area. The advantage of this cell type is that it does not require a steering mechanism for the satellite beam, resulting in lower satellite cost. Please note that this architecture option is already defined for the 5G non-terrestrial networks (NTN) in 3GPP \cite{ref1}, and an illustration is shown in Fig. \ref{fig3}. For our scenario, we make the following assumptions.

\paragraph {Fixed user location on Earth}

We assume a scenario where the UEs have a deterministic location and no mobility is foreseen for them. This allows the BS to obtain the geographical coordinates of the users in the deployment phase, and use this information to help in the radio resource allocation strategy.
	
\paragraph {Perfect pre/post compensation of the common Doppler shift at the GW}

As we have analyzed in our previous work \cite{ref9}, the Doppler shift in a LEO satellite channel can be subdivided into a common Doppler part, experienced by all the users under the same satellite footprint, and a differential Doppler part depending on the relative position of users in the footprint. While the common Doppler shift can be pre/post compensated at the GW since there is enough computational power to continuously estimate and track the position of the satellite, as also proposed in \cite{ref14, ref6}, the same cannot be done at the user side. The low-complexity of the on-ground NB-IoT UEs is of utmost importance. Therefore, assuming an ideal compensation of the common Doppler shift (e.g. the one at the center of the beam) at the GW, leaves out only the differential Doppler shift impacting the transmissions among the BS and the UEs. It is worth reminding here that in the uplink transmission (UE $->$ BS), there exist the possibility of frequency multiplexing of channels. Consequently, the differential Doppler shift among users will cause an overlap among sub-carriers inside the whole NB-IoT carrier, thus introducing the so-called inter-carrier interference (ICI). In contrast, in the downlink transmission, the whole NB-IoT carrier is dedicated to one channel, hence such an overlap does not occur. The differential Doppler part can be estimated by each user in the synchronization phase through the NPSS and NSSS signal. For a more detailed explanation regarding the impact of the differential Doppler shift in the uplink transmission, please refer to \cite{ref9}.
 
		\begin{figure}[!b]
		\centering
		\includegraphics[width = 0.5\textwidth]{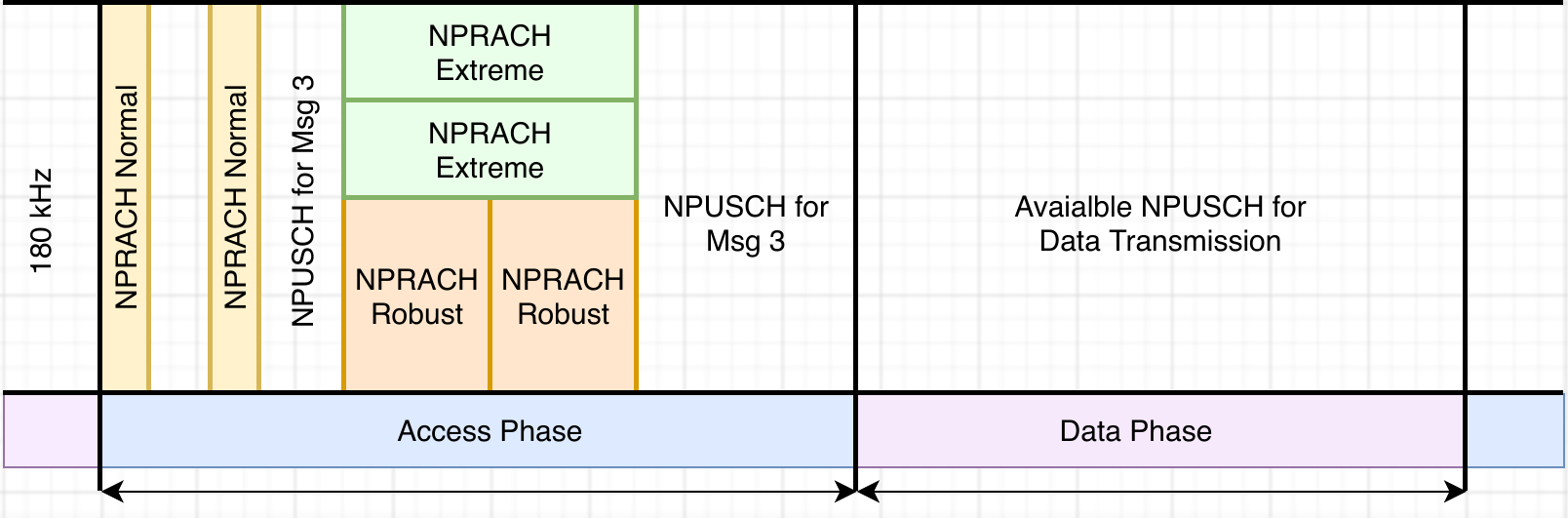}
		\caption{Example of access and data phase configuration in uplink.}
		\label{fig4}
	\end{figure}
\paragraph{Separation of access and data phases}

As we discussed in Section \ref{sec2}, there may be different configurations of the NPRACH channel, which is the one responsible for the access phase. It is possible that the NPRACH occupies all the available frequency resources, or part of them. In addition, the time-duration of the NPRACH channel may be different depending on the targeted coverage class. Of course, the NB-IoT standard does not impose a fixed configuration, rather then give a possible set of parameters under which a certain operator should take into account when designing a system. In our NB-IoT over LEO satellite scenario, it is advantageous to avoid a frequency multiplexing of access and data channels due to the following reasons. First, it enables us to control the differential Doppler shift in the uplink transmission for the data phase, and the corresponding ICI imposed by it, through a resource allocation strategy. The same cannot be done in the access phase because the BS specifies only the NPRACH resources to be utilized, and then users randomly initiate the RA procedure based on their need to transmit data. Of course, overcoming the differential Doppler limit is also an issue to be addressed in the access channel, but it is out of the scope of this work. Second, having distinct operational phases helps in taking proactive "short-term" resource allocation decisions by considering only the latest users with a successful access phase (with the most accurate estimated parameters) to be scheduled in the following data phase. This is crucial for our highly dynamic system where the channel parameters would change quickly over time. In Fig. \ref{fig4} we show an example of a possible access and data phase configuration. Obviously, the time-duration of the access phase and its periodicity would be open for design. Please note that the NPUSCH also plays a role in the access phase since it enables msg 3 transmission where the buffer status is reported. 
			
\section{Radio Resource Allocation Strategy} \label{sec4}

\subsection{Resource allocation objectives}

After the access phase over a certain random access opportunity (RAO), the BS will have to assign resources to users for the data phase. The parameters to be considered are: i) the buffer status reported in msg3 of the RA procedure, ii) the satellite coverage time derived by the UE location on Earth, and iii) the signal to noise ratio (SNR) estimate obtained during the message exchange in the RA procedure. Taking into account the peculiarities of our system, an efficient resource allocation strategy should have the following objectives: 

\paragraph{Maximize the radio resource utilization} The available resources for the data phase are limited, thus there is a need to maximize its utilization. A failure to do so would result in wasted resources and decrease the efficiency of the overall system. Of course, this will also depend on how the access phase is designed in terms of length and periodicity. Since the scope of our work is to focus on the data phase, we consider a fully-loaded system in which there will be sufficient users after the access phase competing for the limited radio resources in the data phase.
	
\paragraph{Select the users to be scheduled} Apart from assigning the time-frequency resources to the users, the BS has to efficiently select the users to transmit in the available radio frames. Certain users will have a shorter satellite coverage time depending on their location in the satellite footprint, hence there is a need that the BS should account for this when assigning resources because these users have a higher probability to be out of the coverage area in the subsequent data phase. Also, giving some priority to the users with better channel conditions would help in increasing the throughput, and consequently more data can be aggregated by the BS. In addition, the users with a higher buffer status may have a more urgent need to offload their data because the buffer size is limited and can lead to missed information due to overflows. A balance among the above-mentioned considerations is needed in order to satisfy the predefined key performance indicators (KPIs) of a certain system. Last but not least, to keep the differential Doppler limit under the allowed threshold, only particular users can be scheduled together in the same frame. 
		
\paragraph{Minimize the message exchange between UEs and the BS} As we already mentioned, in a terrestrial NB-IoT scenario, the BS can decide to empty the buffer of a certain user through subsequent NPUSCH transmissions. This may be advantageous in situations where the available resources are limited in a "short-term", and the data that the user has to transmit can be distributed over a longer time through a set of smaller NPUSCH transmissions. Nevertheless, the same approach cannot be followed for a LEO satellite-based NB-IoT scenario due to the highly dynamic nature of the overall system and the increased RTD in the satellite channel. Notably, every NPUSCH transmission requires the corresponding information regarding the resources to be utilized, which is reported in the NPDCCH. Consequently, the increased message exchanges between the user and the BS would further delay the system because of the considerably higher RTD compared to the terrestrial case. In addition, the estimated parameters under which the resource allocation would be performed may be outdated. For example, assigning the wrong MCS for a certain NPUSCH transmission, for which the channel conditions are worsening over time due to the satellite movement, may result in an erroneous transmission. Therefore, it is highly crucial that the resource allocation strategy should maximize the amount of data encoded in a specific NPUSCH transmission aiming to empty the buffer status of the users in as few attempts as possible. Obviously, there are some boundaries imposed by the NB-IoT protocol which cannot be exceeded, such as the finite number of the MCS or the RUs available.
	
\subsection{Problem Formulation}

Let $N$ denote the set of un-collided devices which perform a successful RA procedure after a specific access phase.After this step the BS will have the SNR estimate of each user $UE_i$, for $i=1, 2, ..., N$, and has to guarantee that the following holds:

\begin{equation}
SNR_i^{rec} \geq SNR(MCS_i, N_i^{sc})
\label{eq1}
\end{equation}
where $SNR_i^{rec}$ is the estimated SNR of the received signal and $SNR(MCS_i, N_i^{sc})$ is the SNR needed to close the link under a specific modulation and coding scheme $MCS_i$ and number of subcarriers $N_i^{sc}$ assigned for uplink data transmission. Please note that according to the NB-IoT standard, $MCS_i$ can take discrete values from 0 to 13 (see Table \ref{tab1}), and the number of subcarriers $N_i^{sc}$ can be 1, 3, 6 or 12. A table that relates these parameters can be found in \cite{ref25}, where the required SNR levels under different $MCS_i$ and $N_i^{sc}$ are obtained. Through a look up table, the serving BS can derive the possible options under which Eq. (\ref{eq1}) would be satisfied.

\begin{figure}[!t]
	\centering
	\includegraphics[width = 0.5\textwidth]{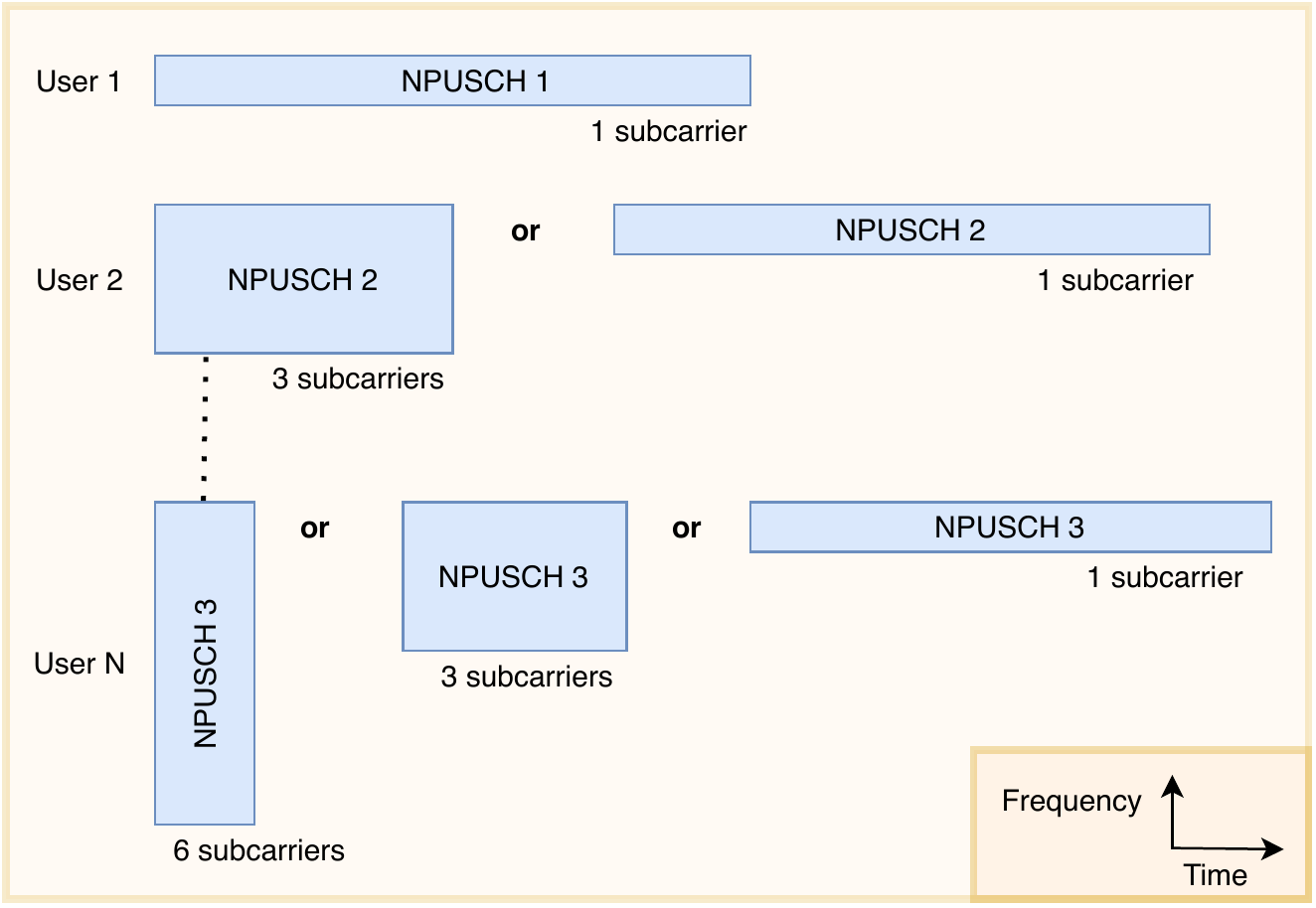}
	\caption{Time-frequency representations of users.}
	\label{fig5}
\end{figure}

Furthermore, after the access phase, each user will have an uplink request with data size $D_i>0$, which it reports in Msg3 of the RA procedure.  Since our goal is to minimize the message exchange between the UE and the BS, it is crucial to empty the buffer size in as few transmissions as possible. Therefore, we need to select the highest reachable MCS for each user. To derive the number of RUs needed for emptying the buffer size $D_i$ of the i-th user, the following must hold:

\begin{equation}
N_i^{RU}=\begin{cases}
\frac{D_i}{16 \cdot R[max(MCS_i)]} , & \text{if $N_i^{sc} = 1 $}\\
\frac{D_i}{24 \cdot R[max(MCS_i)]} , & \text{otherwise}
\end{cases}
\label{eq2}
\end{equation}
where $R[max(MCS_i)]$ is the average data rate of the maximum reachable $MCS_i$ for the i-th user expressed in bits/(subcarrier x slot). It should be noted that according to the standard, the maximum number of RUs $max N_i^{RU} = 10$, so in case a larger number of RUs is needed to empty all the buffer, we chose the maximum one and the user will try to transmit the remaining data on the buffer in another access attempt. The only remaining parameter to be calculated is the duration of the NPUSCH channel $w_i$ as follows:

\begin{equation}
w_i = N_i^{RU} \cdot \frac{N_i^{slot}}{2}(subframes) 
\label{eq3}
\end{equation}
where $N_i^{slot}$ is the number of slots for each RU according to the standard.

\begin{equation}
N_i^{slot}=\begin{cases}
16, & \text{if $N_i^{sc} = 1 $}\\
8, & \text{if $N_i^{sc} = 3 $}\\
4, & \text{if $N_i^{sc} = 6 $}\\
2, & \text{if $N_i^{sc} = 12 $}\\
\end{cases}
\label{eq4}
\end{equation}
Please note that two consecutive slots form one subframe. So basically, after solving equation (\ref{eq1}), (\ref{eq2}), (\ref{eq3}), (\ref{eq4}) the base station will have the NPUSCH representations for all the users with a successful access phase in terms of number of subcarriers in the frequency domain and number of subframes in the time domain. An example is illustrated in Fig. \ref{fig5}. As it can be noted, there may be more than one option in the time-frequency domain for the NPUSCH representation. Obviously, this will depend on the channel conditions. Some users with a successful access phase may close the link only for single-tone transmission, whereas others may have a reliable communication with the BS even for multi-tone. The more subcarriers are utilized for data transmission by the UE, the more difficult it is to close the link because the same power of the UE will be distributed over a larger bandwidth \cite{ref18}. Notably, this results in shorter channels in time due to Eq. (\ref{eq4}). So far, we have addressed only one of the resource allocation objectives, which was to minimize the message exchange among UE and BS, through selecting the highest reachable MCS and number of RUs depending on the user channel conditions and buffer status report.

The problem to be solved now is how to efficiently select the scheduled user set which optimize the available resource utilization in the data phase. This can be modeled as a packing problem, since we have to select a set of rectangles (users), which best fit in another rectangle (available resources for NPUSCH transmission), and maximizes a certain profit function (to be designed). Our specific problem falls into the 0-1 Two-dimensional Knapsack Problem (KP). It is a 0-1 KP because we have to select or reject a user from scheduling in the current data phase, and 2D-KP because we have to assign the exact time-frequency resources for each scheduled user. Since we have discrete values for the frequency (subcarriers) and time domain (subframes), we can formulate the problem as an integer linear programing (ILP) model as follows. 

Let $P_i$ be the profit of selecting the $i$-th user for scheduling in the data phase, $X = \{p \in Z | 1 \leq p \leq W \}$ represent the time domain and $Y = \{q \in Z | 1 \leq q \leq B \}$  the frequency domain. By defining $z_{ipq}$ and $g_{ij}$ as follows:

\begin{equation}
z_{ipq}=\begin{cases}
1, & \text{if $UE_i$ is selected for scheduling with} \\ 
	& \text{its left bottom corner (p,q)}\\
0, & \text{if $UE_i$ is not selected for scheduling}\\
\end{cases}
\end{equation} 
\begin{equation}
g_{ij}=\begin{cases}
0, & \text{if $UE_i$ belongs to group $G_j$}\\
1, & \text{otherwise}\\
\end{cases}
\end{equation} 
we are able to mathematically formulate our problem through equation (7) - (13). 
\begin{table*}[!t]
		\normalsize
	
	\setcounter{equation}{6}

	\begin{align}
	&\max_Z \text{\vspace{1cm}} \sum_{i=1}^N \sum_{p \in X} \sum_{q \in Y} P_i \cdot z_{ipq}\\
	&\text{subject to} \notag \\
	&\sum_{i=1}^N \sum_{\{p \in X| r - w_i + 1 \leq p \leq r \}} \sum_{q \in Y} b_i \cdot z_{ipq} \leq B \hspace{2.52cm} \forall  r\in X \\
	& \sum_{i=1}^N \sum_{p \in X} \sum_{\{q \in Y| s - b_i + 1 \leq q \leq s \}} w_i \cdot z_{ipq} \leq W \hspace{2.4cm} \forall s\in Y \\
	& \sum_{i=1}^N \sum_{\{p \in X| r - w_i + 1 \leq p \leq r \}} \sum_{\{q \in Y| s - b_i + 1 \leq q \leq s \}} z_{ipq} \leq 1 \hspace{1.1cm}
	 \forall r\in X, \forall s\in Y \\
	& \sum_{p \in X}\sum_{q \in Y} z_{ipq} \leq 1 \text{\hspace{5.65cm} for i = 1,2, ... N} \\
	& z_{ipq} \in \{0, 1\} \text{\hspace{6.3cm} for i = 1,2, ... N},\text{\,} p \in X,\text{\,} q \in Y \\
	& \sum_{j=1}^M \sum_{i=1}^N \sum_{\{p \in X| r - w_i + 1 \leq p \leq r \}} \sum_{q \in Y} g_{ij} \cdot z_{ipq} = 0 \hspace{1.85cm} \forall r\in X
	\end{align}
	\hrulefill
\end{table*} 
	\begin{figure}[!t]
	\centering
	\includegraphics[width = 0.4\textwidth]{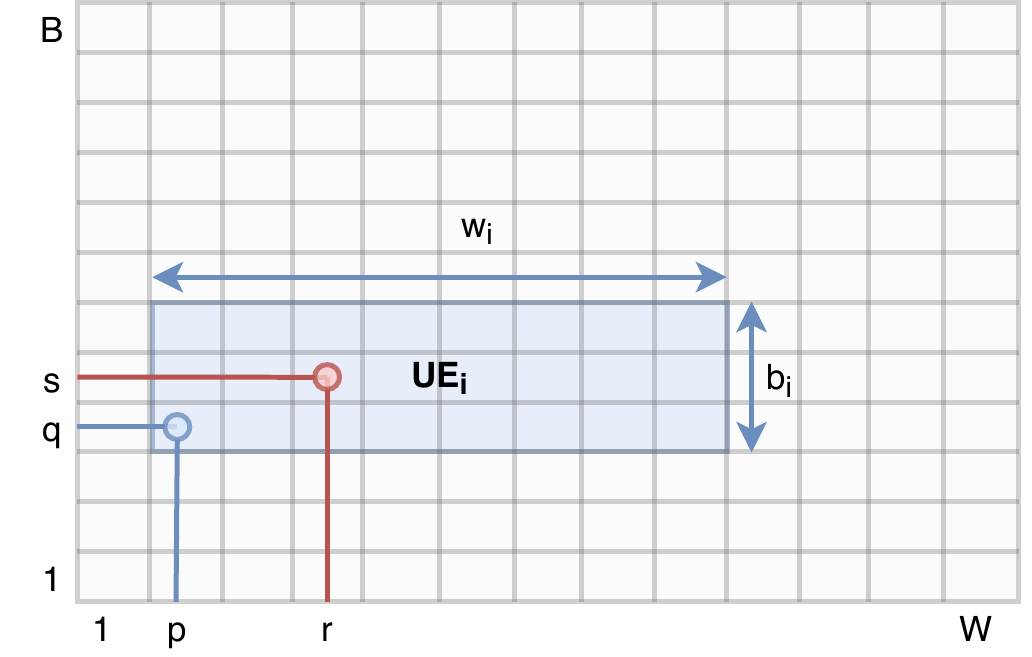}
	\caption{Scheduling a user in the available resources.}
	\label{fig6}
\end{figure}
Please note that $B$ is the available bandwidth expressed in subcarriers (12 in our NB-IoT scenario) and $W$ the available subframes for the data phase. Fig. \ref{fig6} may be helpful to understand the other parameters of equation (7) - (13). The objective function (7) represents a scheduling pattern of maximum profit. Constraints (8) ensures that for any subframe, the sum of the assigned bandwidths $b_i$ of the scheduled users (utilizing the particular subframe) does not exceed the maximum bandwidth available B. In analogy, constraint (9) ensures the same, but in the other dimension. The non-overlapping constraints are given in (10), imposing any point (r, s) of the time-frequency grid is utilized by one and only one user. Constraint (11) is related to the fact that each user can be scheduled no more than once. The variables domain is given by (12). Last but not least, constraint (13) ensures that at any subframe only users belonging to the same group can be scheduled together. The groups are created in such a way that users inside the same group $G_j$ do not violate the differential Doppler limit. The higher the distance among users along the x-axis (corresponding to the satellite movement direction), the higher the differential Doppler would be, whereas the contribution of the y-axes is negligible \cite{ref9}. Depending on the configurations of a particular system (carrier frequency, satellite altitude etc.), the maximum tolerated distance along x-axes can be calculated. This allows us to create fixed groups of users in the satellite footprint as demonstrated in Fig. \ref{fig7}.
The problem to be solved (7) - (13) is strongly NP hard and thus any dynamic programming approach would result in strictly exponential time bounds. It is crucial to further simplify the problem.

\begin{figure}[!b]
	\centering
	\includegraphics[width = 0.38\textwidth]{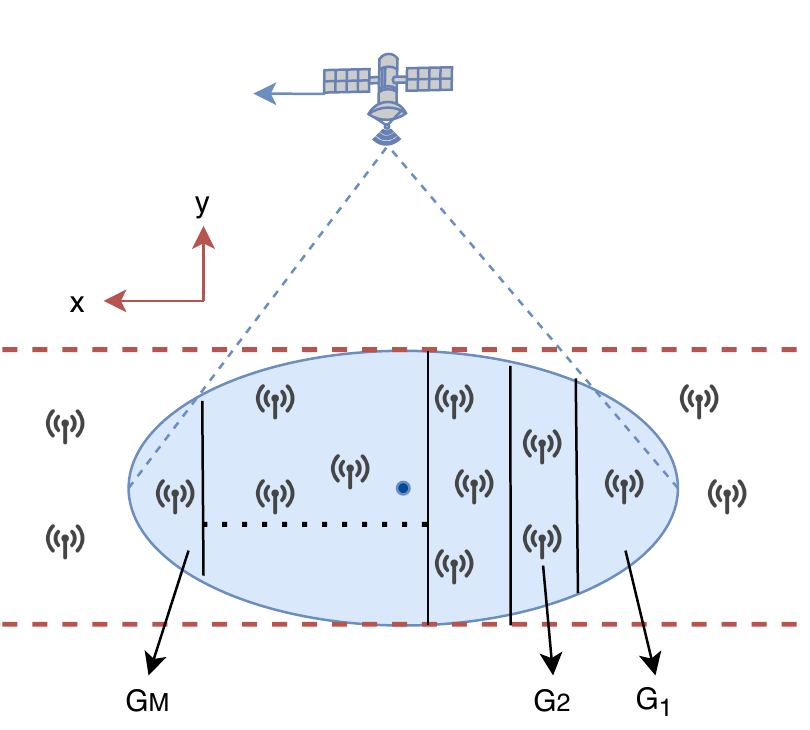}
	\caption{Satellite footprint divided in smaller regions.}
	\label{fig7}
\end{figure}

\subsection{Problem Re-formulation}

The problem can be relaxed by removing the last constraint (13) and solving (7) - (12) for each of the user groups $G_j$ separately. Of course, this would require an optimal time resource share $W_j$ among different groups, obtained by solving the following problem. Let

\begin{equation}
S_j = \sum_{i \in G_j} P_i
\end{equation}
be the profit for the $j$-th group created in the satellite footprint given by the sum of the profits of every user inside that particular group. Then find:
\begin{align}
&\max_{w_j} \min_j \frac{W_j}{S_j}\\
&\text{subject to} \notag \\
& \sum_{j=1}^M W_j \leq W \\
& W_j \in \mathbb{N} \hspace{1.2cm} \forall j \in {1,2,...,M}
\end{align}
where $M$ is the number of created user groups. This can be quickly solved through linear programing. The remaining problem (7) - (12) would be a typical 2D-KP where the items are not allowed to rotate.  The main algorithms in the literature that find the exact solutions for this type of problems can be found in \cite{ref26, ref27, ref28}. However, the proposed solutions can quickly converge only for small number of items. The largest problem they solve is for $N=50$. It is worth highlighting here that in our scheduling problem the number of users to be scheduled competing for the available resources can be higher than $50$, even in the case or smaller groups. In addition, for every user there may be more than one representation, significantly increasing the search space, and consequently the complexity of the problem.  Therefore, it is essential to reduce the complexity of the problem as much as possible intending to accommodate the high dynamicity of the system. 

To further relax the problem, it is possible to fix the transmission mode only to single-tone. In other words, this would mean that the UEs can transmit only utilizing 1 subcarrier in the frequency domain. This approach is also advantageous from the link-budget perspective since the available power at the user is distributed over a narrower bandwidth \cite{ref25}. Fixing the allowed transmission mode only to single-tone would transform our 0-1 2D-KP into a 0-1 Multiple Knapsack Problem (MKP). This is a 1D problem because the only parameter deciding the item representation is the time-length while the frequency is fixed, and it is a multiple knapsack because there will be $B$ subcarriers (or knapsacks) where users can be scheduled. The problem to be solved would be how to optimally select $B$ disjoint subset of users to be scheduled in the $B$ available subcarriers without exceeding the knapsack capacities (available time for data phase). Mathematically it can be written:
\begin{align}
&\max_z \text{\vspace{1cm}} \sum_{i=1}^{B} \sum_{k=1}^{N_j} P_{k} \cdot z_{ik}\\
&\text{subject to} \notag \\
& \sum_{k=1}^N w_k \cdot z_{ik} \leq W_j \text{\hspace{1cm} for \hspace{0.1cm}} i = 1,2,...,B \\
&\sum_{i=1}^{B} z_{ik} \leq 1 \text{\hspace{2cm} for \hspace{0.1cm}}  k = 1,2,...,N_j \\
& z_{ik} \in \{0, 1\} \text{\hspace{2cm} for \hspace{0.1cm}} i = 1,2, ... B, \\
& \hspace{4.4cm} k = 1,2, ... ,N_j \nonumber
\end{align}
where
\begin{equation}
z_{ik}=\begin{cases}
1, & \text{if $k$-th UE is selected to be scheduled }\\
	& \text{in the $i$-th subcarrier}\\
0, & \text{otherwise}\\
\end{cases}
\end{equation}
$N_j$ and $W_j$ are the number of users and the available time for data phase of the $j$-th user group. Evidently, the problem has to be solved for each of the user groups (see Fig. \ref{fig7}). The re-formulated problem (18) - (21) is less complex because we have fixed one of the dimensions and reduced the search space. Every user to be scheduled will have only one representation (instead of many as depicted in Fig. \ref{fig5}) given only by its NPUSCH time-length.

\section{Solutions to the resource allocation problem} \label{sec5}

\subsection{Exact Solution}

For the exact solution of the 0-1 MKP, we refer to the method proposed in \cite{ref29} where an exact algorithm is presented, particularly designed for solving large problem instances. In general terms, this algorithms utilizes three main steps shown below:

\begin{itemize}
\item \textbf{Derive an upper bound:} The algorithm utilizes the surrogate relaxation to derive an upper bound, through relaxing constraint (20). Doing so, the 0-1 MKP problem will be converted in an ordinary 0-1 KP, easily solvable through dynamic linear programming:
\begin{align}
&\max_{z^\prime} \text{\vspace{1cm}}  \sum_{j=1}^{N_j} P_{k} \cdot z_{k}^{\prime}\\
&\text{subject to} \notag \\
& \sum_{k=1}^{N_j} w_k \cdot z_{k}^{\prime} \leq B \cdot W_j \\
& z_{k}^{\prime} \in \{0, 1\} \text{\hspace{2cm} for } k = 1,2, ... ,N_j
\end{align}
where the introduced decision variable $z_k^{\prime}= \sum_{i=1}^{B}z_{ik}$ indicates whether the $k$-th user is chosen for scheduling in any of the subcarriers (knapsacks) and the new knapsack capacity (data phase time-length) $W^{\prime} = B \cdot W_j$ is given by the capacity of the united knapsacks. 

\item D\textbf{erive a lower bound:} The lower bound is obtained by splitting the chosen users after equation (23) - (25) for the united knapsack into the $B$ individual knapsacks.
This is done by solving a series of Subset-sum Problems as follows:
\begin{align}
&\max_{z} \text{\vspace{1cm}}  \sum_{k=1}^{N^{\prime}_{k}} w_{k} \cdot z_{k}\\
&\text{subject to} \notag \\
& \sum_{k=1}^N w_k \cdot z_{k} \leq  W_j \\
& z_{j} \in \{0, 1\} \text{\hspace{2.6cm} for}  j = 1,2, ... ,N^{\prime}_j
\end{align}
where $N^{\prime}_j$ is the number of users in the subset (not to be confused with $N_j$) from the optimal solution $z^{\prime}$ of problem (23) - (25). It can be noted that this problem tries to fit as many items (from the ones selected by the upper bound solution) as possible in every individual knapsack.

\item \textbf{Bound and bound (B\&B) algorithm:} This algorithm is a particular tree-search technique which utilizes the lower bound calculation to determine the branches to follow in the decision tree. If at any tree node, through the lower bound calculation procedure, we are able to fit all the selected items (after solving upper bound) into the $B$ knapsacks, the lower bound equals the upper bound, and we may immediately backtrack. There is no need to explore other tree nodes because the optimal solution is found. Otherwise a feasible solution is obtained which contains some (but not all) of the items, and the algorithm continues by exploring other tree nodes considering other items. 
\end{itemize}

Notably, how fast this algorithm converges to the optimal solution will depend on the particular problem instance, and the quality of the obtained upper bound. Taking into account equation (\ref{eq3}) and (\ref{eq4}), and the fact that we fix the transmission mode only to single-tone, we can observe that the possible time-length of the users $w_k$ to be scheduled will be:
\begin{equation}
w_k = 8 \cdot N_k^{RU} (subframes)
\end{equation} 
Regardless of the users we select to schedule, their merged time-duration will be a multiple of 8. Having knapsack capacities $W_j$ that are not a multiple of 8, will lead to radio resource wastage and a week upper bound calculation (slowing down the convergence of the B\&B algorithm). As a result, it is crucial for our scenario to take advantage of this feature and impose it in equation (15) - (17), where we obtain the available time shares ($W_j$) among different groups. Changing constraint (17) to:
\begin{equation}
 W_j \in 8 \mathbb{N} \hspace{1.2cm} \forall j \in {1,2,...,M}
\end{equation}
will not only avoid resource wastage, but also significantly improve the upper bound calculation in (23) - (25), resulting in a faster convergence of the B\&B algorithm.

\subsection{Greedy Algorithm}

This is a quick procedure to find a feasible solution as proposed in \cite{ref31}. The first step is to order the items according to their profit per unit weight:
\begin{equation}
\frac{P_1}{w_1} \geq \frac{P_2}{w_2}\geq...\geq \frac{P_{N_j}}{w_{N_j}}
\end{equation}
Then the items are consecutively inserted into the knapsack until the first item, $s$, is found which does not fit. This is called the critical item. 
\begin{equation}
s = min \Big\{l: \sum_{k=1}^l w_k > W_j \Big\}
\end{equation}
The greedy algorithm is firstly applied to the first knapsack, then to the second one by using only the remaining items, and so on. Through this procedure, the overall profit $G$ of the selected users to schedule would be:
\begin{equation}
G = \sum_{i=1}^{B} \sum_{k=s_{i-1}}^{s_i -1} P_k
\end{equation} 
Obviously, the greedy algorithm will most probably waste resources. When a critical item is found, the remaining capacity of the i-th knapsack $\overline{W_j}$ (remaining available time for users to schedule in the i-th subcarrier) will be empty. 
\begin{equation}
\overline{W_j} = W_j - \sum_{k=s_{i-1}}^{s_i - 1} w_k
\end{equation}
However, this is a good algorithm for maintaining very low complexity since it moves only forward (index $i$ and $k$ in equation (33) always increasing ). 

\subsection{Approximate Solution}

Exploiting also the unique features of our scenario, it is possible to further improve the solution of the greedy algorithm. The one that we propose here consist in four main steps:

\begin{itemize}
	
\item \textbf{Step 1:} Order the items according to their profit per unit weight as in equation (31).

\item \textbf{Step 2:} Merge the $B$ knapsacks into 1 with single capacity $W^{\prime} =B \cdot W_j$ and solve it through the Greedy algorithm. The advantage of this approach is that it reduces the resource wastage compared to the case where you have to apply the Greedy algorithm for each individual knapsack.

\item \textbf{Step 3:} Re-sort the selected items according to their weight $w$, in a decreasing order. This is important since we want to place in each knapsack the "bad" items first with the highest weights $w$, in order to leave the "good" items (with low weight) at the end. Please note that the more items we place in the knapsack, the smaller the remaining capacity would be. Therefore, getting rid of the items with high weight first while the capacity is still large will increase the probability of fitting the remaining items with lower weights into each individual knapsack.

\item \textbf{Step 4:} Place the items into the knapsacks according to the order in Step 3, and at every iteration select the knapsack with the maximum remained capacity $\overline{W_j}$.

\end{itemize}
Notably, this algorithm tries to imitate the steps of the exact solution, but with reduced complexity, since we derive an upper bound (23) - (25) through the Greedy algorithm, and then attempt to separate the selected items of the upper bound solution into $B$ individual knapsacks through heuristics. Specifically, the complexity of the approximate solution is $O(n\,log\,n)$ for sorting plus $O(n)$ for placing the items into the knapsacks. Although this represents the same complexity as the Greedy algorithm, the running time would be higher because the sorting procedure and the item placement is repeated more than once.
\begin{table}[!b]%
	\caption{Simulation Parameters for algorithm comparison.\label{tab2}}
	\centering
	
	\begin{tabular}{|c|c|}
		\hline
		\textbf{Parameter} & \textbf{Value}  \\
		
		\hline
		
		\textbf{User Profit $P_k$} & Integer values randomly selected   \\
		\textbf{} & inside the range [1 1000]   \\
		\hline
		\textbf{User Weights $w_k$}  & \{8, 16, 24, 32, 40, 48, 64, 80\} SF  \\ 
		\textbf{expressed in SF} & According to Table \ref{tab1} and Eq. (\ref{eq3}) and (\ref{eq4})\\ \hline
		\textbf{Number of Knapsacks $B$} & 12 (subcarriers) \\
		\textbf{fixed for NB-IoT} &  \\ \hline
		\textbf{Knapsack Capacity $W_j$} & 160 SF \\ 
		\textbf{expressed in SF} &  \\ \hline
		\textbf{Number of Users $N$} & [150 500] \\
		\textbf{to be scheduled} &  \\ \hline
		\textbf{Number of Iterations} & 1000\\
		\textbf{(problem instances)} & \\
		\hline
	\end{tabular}
\end{table}
\begin{figure}[!t]
	\centering
	\includegraphics[width = 0.45\textwidth]{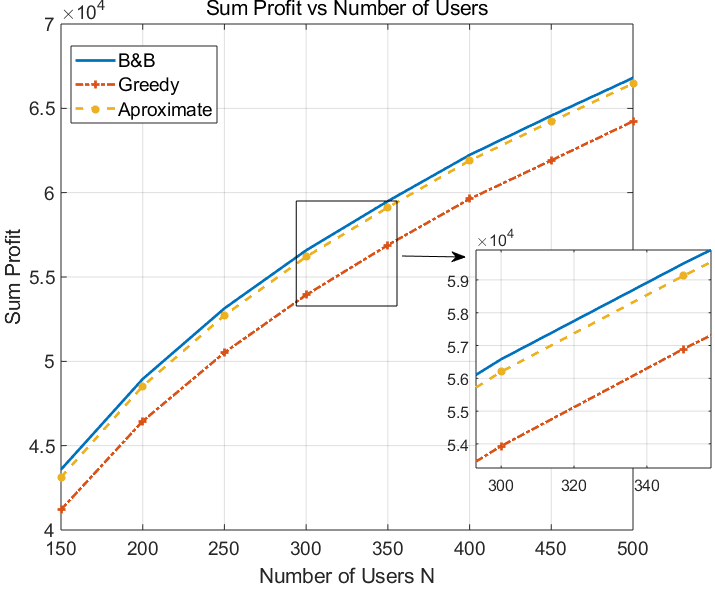}
	\caption{Average sum profit under different number of users.}
	\label{fig8}
\end{figure}
\begin{figure}[!t]
	\centering
	\includegraphics[width = 0.48\textwidth]{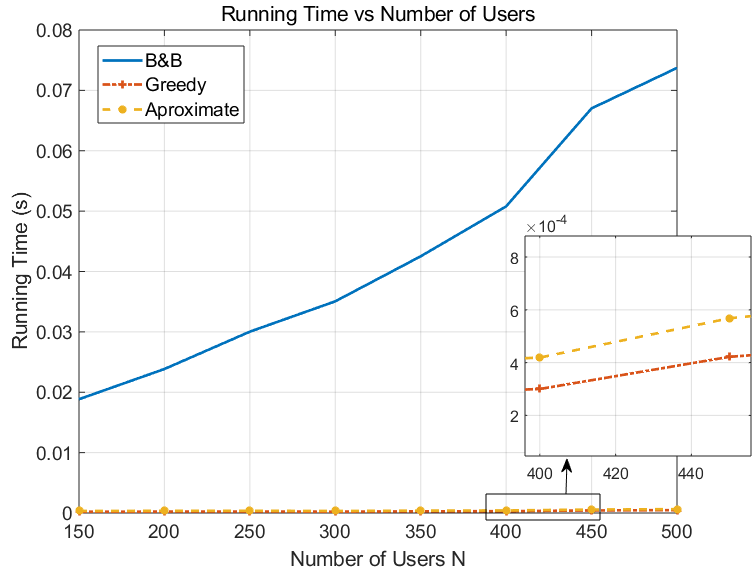}
	\caption{Average running time under different number of users.}
	\label{fig9}
\end{figure}
\subsection{Comparison among algorithms}

To compare among the algorithms, we solve our 0-1 MKP problem (18) - (21) in a hypothetical scenario where many users compete for scheduling in the available data phase. For the numerical simulations, we use the parameters summarized in Table \ref{tab2}. Please note that we run the algorithm for different problem instances and average over the whole number of iterations with the aim of obtaining the average sum profit and the average running time required by the three different algorithms, as demonstrated in Fig. \ref{fig8} and \ref{fig9}. Also, we make sure to select a knapsack capacity $W_j$ large enough to avoid user exclusion, and the number of users $N$ large enough to avoid trivial solutions where all the users can be scheduled together. Last but not least, we assume the candidate users with a successful access phase belong to the j-th group, thus the differential Doppler limit inside the radio SF is not violated.

As it can be noted, the optimal solution is given by the B\&B algorithm, through which the maximum sum profit is obtained. Nevertheless, its higher complexity requires more time to converge to the optimal solution compared to the other investigated algorithms in this paper. The Greedy algorithms gives the worst results in terms of sum profit, but the best ones in terms of running time. Our proposed Approximate solution performs very close to the exact one in terms of sum profit, and very close to the Greedy one in terms of running time. This is due to the fact that the knapsack capacities $\overline{W_j}$ and the user weights $w_k$ are correlated (both multiple of 8). For other problem instances, with very random capacities and weights, the performance of this algorithm would be comparable to the Greedy one.

	\begin{figure}[!t]
	\centering
	\includegraphics[width = 0.5\textwidth]{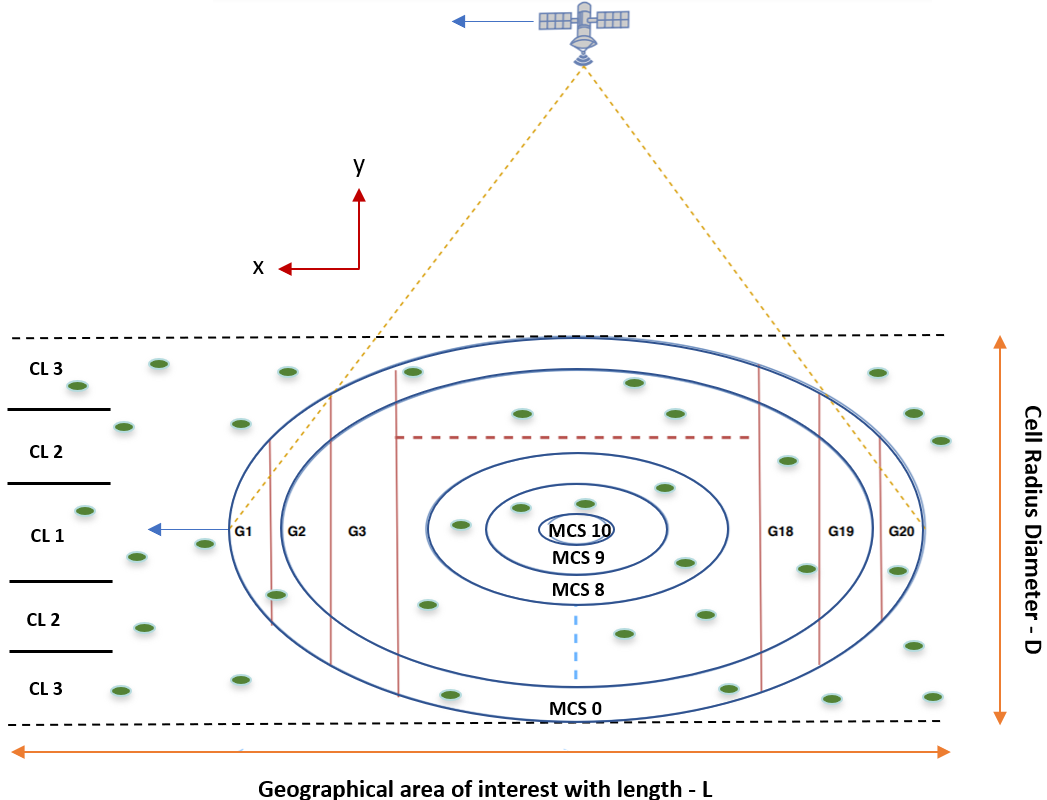}
	\caption{Reference Scenario for Numerical Simulations.}
	\label{fig10}
\end{figure}

\section{Numerical Results} \label{sec6}

Using the best possible algorithm for the resource allocation problem, as shown in the previous section, does not guarantee the efficiency of the overall NB-IoT system, which will greatly depend on the way we assign priorities (profits) to various users. In addition, it is crucial to define specific key performance indicators (KPIs) for our NB-IoT via LEO satellite system, as a means to find optimal operational points in the profit function design and compare among different algorithms.

\subsection{Simulation Setup}

To perform simulations, we consider a dynamic scenario where a LEO satellite with altitude $h$ is moving over a specific geographical area of interest with length $L$ and width $D$. The latter corresponds also to the diameter of the satellite footprint (cell) ob ground (please refer to Fig. \ref{fig10}). The users are uniformly distributed over the geographical area of interest with coordinates (x;y). Based on their specific locations, every user will have a certain overall coverage time by the satellite $t_{sat}$. In addition, the satellite altitude $h$ will also determine its periodicity over the geographical area of interest. Regarding the traffic model, we use the one defined in the 3GPP for the IoT applications \cite{ref30}. Users will generate packets following a Pareto distribution with shaping parameter $\alpha=2.5$ range between 20 and 200 Bytes. The periodic inter-arrival time of new generated packets is: 1 day (40 \%), 2 hours (40 \%) , 1 hour (15 \%), and 30 minutes (5 \%).

\subsection{Profit Function Definition}

To assign profits to different users, the BS has to be based on the estimated/known parameters. As we stressed out in Section \ref{sec4}, these parameters include a) the buffer status reported msg3 of the RA procedure, b) the satellite coverage time derived by the user location of the users on-ground and c) the SNR estimate obtained during the message exchange in the access phase. The SNR estimate of the signals transmitted by various users will greately depend by the satellite antenna parameters and the antenna beam pattern, whereas the ones of the NB-IoT devices will be the same as in the terrestrial case. To simplify our simulator, we assume that the antenna at the satellite is designed in such a way that allows transmission at MCS 0 for the users in the edge of the satellite footprint, and at MCS 10 (which is the maximum reachable MCS for single-tone transmission) for the users in the center of the footprint. Two contributors towards this SNR change among users at the satellite beam edge and center are the difference in the propagation path loss and satellite antenna gain. While we can calculate the former one, the latter would depend on how the satellite antenna is designed. However, this assumption does not impact our analysis, since our goal is to find a simple way in representing a fair (reasonable) SNR change among users on ground for our simulator. Through this assumption, it is possible to obtain the maximum reachable MCS for every user just by utilizing their specific coordinates in the geographical area of interest. In a realistic scenario, the BS would have to rely on the SNR estimate in the access phase. Overall, the profit function would have the following form:
\begin{equation}
P_k = \alpha \frac{D_k}{max(D)} + \beta \frac{1 + MCS_k}{ 1+ max(MCS)} + \gamma  (1 - \frac{t_k^{sat}}{max(t_k^{sat})})
\end{equation}
where the first term driven by the parameter $\alpha$ assigns priority to the users with higher buffer status $D_k$, the second driven by $\beta$ to the ones with higher SNR values (and consequently $MCS_k$), and the third one driven by $\gamma$ to the users with lower satellite coverage time $t_k^{sat}$. Obviously, the closer are the users to the edges of the geographical area of interest, the smaller would be the satellite coverage time, thus is crucial to take this into account. Furthermore, users with better channel conditions guarantee a higher throughput. Whereas, the advantage of scheduling users with higher buffer status is to avoid data expiration. Varying $\alpha$, $\beta$ and $\gamma$ will result in different KPIs, which we will define in the following section. 

\subsection{KPI Definition}

The first important KPI to evaluate our system is the achievable throughput for the data phase given by:
\begin{equation}
R = \frac{T_{bits}}{n \cdot W \cdot m}
\end{equation}
where $W$ is the available time (converted in ms, 1 SF = 1 ms) between two access phases for data transmission, $n$ is the number of data phases during one satellite passage over the geographical area of interest and $m$ is the number of satellite passages. The total transmitted bits during this time is given by $T_{bits}$. In addition to the throughput, it is important to guarantee a fair distribution of service among users in the geographical area of interest. To do so, we define the second KPI as follows. Let $u_k = s_k/d_k$ be the satisfaction ratio per user $k$, where $s_k$ represents the sent bits by the k-th user in the overall data phase opportunity and $d_k$ is the demand in bits (sum of packets generated with time). Then, to compute the user fairness, we utilize the Jain's index give by:
\begin{equation}
J = \frac{(\sum_{k=1}^{N} u_k)^2}{N \cdot \sum_{k=1}^{N}u_k^2}
\end{equation}
where $N$ is the number of users in the geographical area of interest. In addition to the user fairness given by equation (41), there is a need for a third KPI, which gives an indication on the fairness among different coverage levels. Obviously, in our area of interest, there will be users with always bad coverage conditions, which are the most difficult ones to serve. A high user fairness $J$ does not guarantee that the service is fairly distributed among various coverage levels. We define three coverage levels (CL), in analogy to the terrestrial NB-IoT case. Users with good coverage condition (CL 1) will be the ones reaching MCS 7 - 10, medium coverage (CL 2) reaching MCS 3-6, and bad coverage (CL 3) reaching MCS 0-2 (refer to Figure \ref{fig10}). We utilize again the Jen's formula for measuring the fairness in different CL expressed as:
\begin{equation}
J_{CL} = \frac{(\sum_{i=1}^{N_{CL}} U_{i})^2}{N_{CL} \cdot \sum_{i=1}^{N_{CL}}U_i^2}
\end{equation}
where $N_{CL}$ represents the number of the CL, $U_i = (\sum_{k=1}^{N_i} s_k)/(\sum_{k=1}^{N_i} d_k)$ is the group satisfaction given by the ratio of the total transmitted bits with the total generated bits from the users in a certain CL. Please note that $N_i$ is the number of users inside the i-th CL.
\begin{table}[t]%
	\caption{Simulation Parameters for system level analysis.\label{tab3}}
	\centering
	
	\begin{tabular}{|c|c|}
		\hline
		\textbf{Parameter} & \textbf{Value}  \\
		
		\hline
		
		\textbf{Geographical area length $L$} & 4000 km  \\
		\hline
		\textbf{Satellite beam diameter $D$}  & 400 km \\ 
		\hline
		\textbf{Satellite altitude $h$} & 1000 km \\
		\hline
		\textbf{NB-IoT Carrier Frequency} & 2 GHz \\ 
		\hline
		\textbf{Data phase time $W$} & 3600 SF (ms) \\
		\hline
		\textbf{Number of satellite passages $m$} & 480\\
		\hline
		\textbf{Number UEs in the geographical area $N$} & 600 000\\
		\hline
		
	\end{tabular}
\end{table}
\begin{figure}[!t]
	\centering
	\includegraphics[width = 0.44\textwidth]{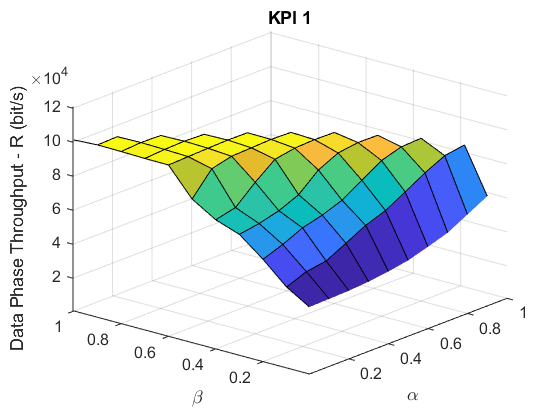}
	\caption{Data phase throughput - maximum at $\alpha = 0.2$, $\beta = 0.7$, $\gamma = 0.1$. \textcolor{blue}{(Note: $\alpha + \beta +\gamma = 1$).}}
	\label{fig12}	
\end{figure}
\begin{figure}[!t]
	\centering
	\includegraphics[width = 0.44\textwidth]{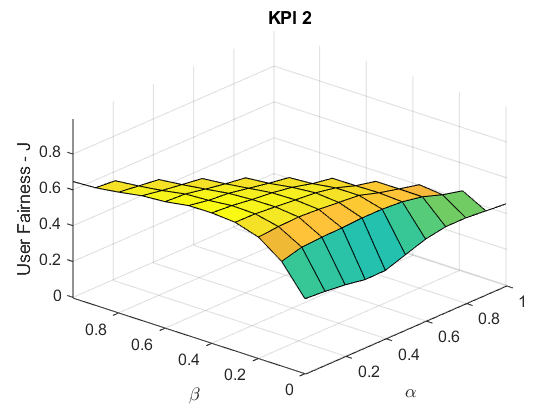}
	\caption{Jen's index for user satisfaction - maximum at $\alpha = 0$, $\beta = 0.4$, $\gamma = 0.6$. \textcolor{blue}{(Note: $\alpha + \beta +\gamma = 1$).}}
	\label{fig13}	
\end{figure}
\begin{figure}[!t]
	\centering
	\includegraphics[width = 0.44\textwidth]{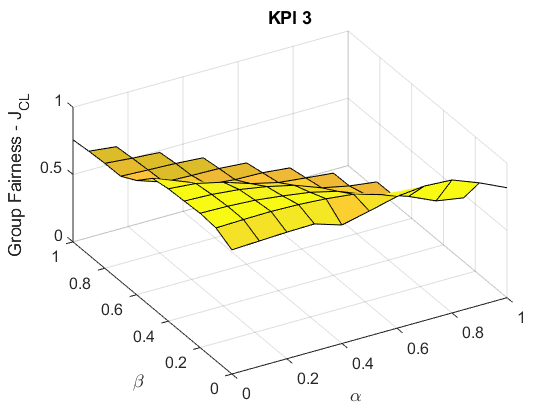}
	\caption{Jen's index for coverage level satisfaction - maximum at  $\alpha = 0.8$, $\beta = 0$, $\gamma = 0.2$. \textcolor{blue}{(Note: $\alpha + \beta +\gamma = 1$).}}
	\label{fig14}	
\end{figure}
\subsection{Performance trade-offs evaluation}

To evaluate the performance of our system, we perform numerical simulations in MATLAB using the simulation parameters summarized in Table \ref{tab3}. Please note that for the selected carrier frequency of 2 GHz, the maximum distance along x-axis that does not violate the differential Doppler limit is 20 km \cite{ref9}. This means that there will be 20 user groups needed for our scenario. Selecting a higher frequency would result in a bigger number of user groups and vice versa.

\begin{figure}[!t]
	\centering
	\includegraphics[width = 0.49\textwidth]{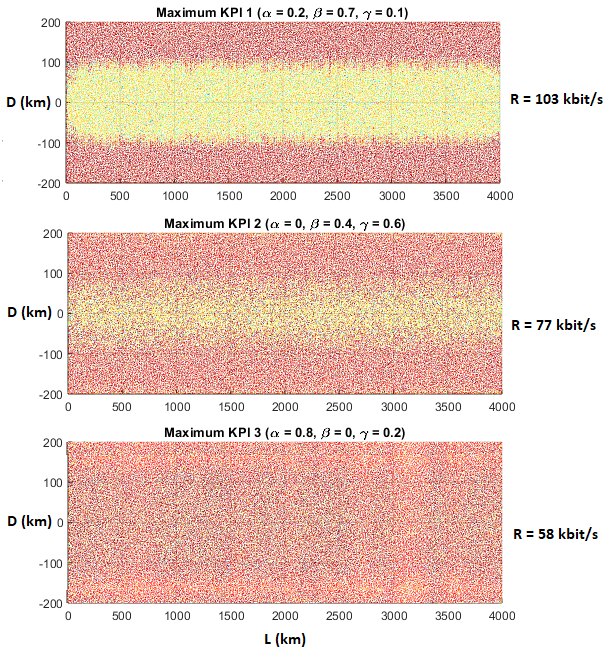}
	\caption{Color mapping of remaining bits in the users' buffer.}
	\label{fig15}	
\end{figure}

The simulation results are shown in Fig. \ref{fig12}, \ref{fig13} and \ref{fig14}. As it can be noted, maximizing the data phase throughput is mainly driven by the parameter $\beta$. This is because scheduling users with better channel conditions for the data phase directly increases the throughput. Whereas, for the user fairness, assigning a higher importance to $\gamma$ allows for a better distribution of services towards users with lower satellite coverage time. Finally, for the third KPI, the maximum fairness among different coverage levels is  reached when we assign most of the profit to the users with a high buffer status. To better understand these results, in Fig. \ref{fig15} we show through a color mapping the remaining data in the users' buffer after the 480 satellite passages at maximum KPIs. Please note that the higher remaining data in the buffer the darker the color. Selecting which KPI to optimize will greatly depend on the parameters of a realistic system, which in our case are unknown, such as, the number of LEO satellites in a constellation, the number of NB-IoT carriers to offer service over an area, the user distribution, and the satellite antenna characteristics. Nevertheless, our hypothetical scenario allows us to make the following analysis.

\paragraph{Maximizing KPI 1}

Maximizing the overall throughput of the system, which is our case it reaches 103 kbit/s, would result in a narrower beam mainly serving user that are closer to the satellite (higher antenna gains and lower propagation path losses). More specifically, in our simulated scenario, we notice from Fig. \ref{fig15} that to satisfy all the user demand we need to double the amount of satellites so as to cover with service the rest of the unsatisfied users placed at the edge of the geographical area. Obviously, this is only an observation and the specific number of satellites required would depend on a realistic user demand and satellite antenna characteristics. Nevertheless, we can conclude that maximizing KPI 1 would be favourable for a system with a large number of LEO satellites in a constellation with highly directive antennas.

\paragraph{Maximizing KPI 3}

Maximizing KPI 3 gives us the best distribution of services among users in different coverage levels, but the lowest throughput (58 kbit/s). Depending on the data demand over a particular area, this scenario would be beneficial for operators targeting to aggregate the IoT traffic through a lower number of LEO satellites with a wide antenna beam on Earth. Nevertheless, to satisfy the data demand and fully serve a particular area, more than one NB-IoT carrier may be required. Explicitly, for our simulated scenario, we would need to double the number of NB-IoT carriers since the resulting throughput is approximately halved.
\paragraph{Maximizing KPI 2}

The maximum user fairness provides a better distribution of services compared to the case with maximum KPI 1, but it fails to guarantee a fair distribution over users in different coverage levels. It still gives a certain priority to users with good channel conditions because it allows more data to be aggregated by the satellite, thus quickly increasing the number of users satisfied with service (and consequently the user fairness). Maximizing KPI 2 would be advantageous in a system where you want to save spectrum (NB-IoT carriers) and at the same time keep the number of satellites in a constellation low, as in the case of maximum KPI 3. 

Another interesting fact worth mentioning here is that a certain system may be in favour of different performance targets at different times. Obviously, in our hypothetical scenario we consider an overloaded system where users are uniformly distributed in order to stress test our algorithm. However, in a realistic scenario, the user distribution will change over time, depending weather the satellite is passing over a city (high number of congested users) or an ocean (e.g. users in ships distributed over a larger geographical area). 

\begin{figure}[!t]
	\centering
	\includegraphics[width = 0.48\textwidth]{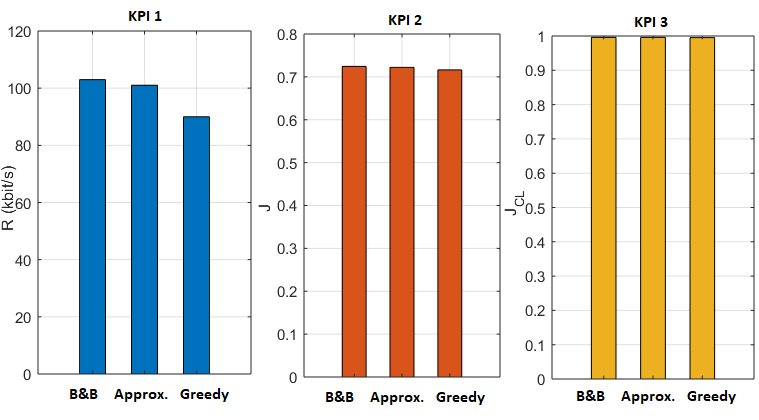}
	\caption{Comparison among algorithms for maximum values of KPI 1, KPI 2 and KPI 3.}
	\label{fig16}	
\end{figure}

The simulation results shown so far where obtained by utilizing the optimal solution of the 0-1 MKP problem. We repeat the simulation also for the Greedy and the Approximate algorithm proposed in this paper, and the comparison results are shown in Fig. \ref{fig16}. As it can be noted, the proposed Approximate algorithm achieves a maximum throughput of 101 kbit/s, being very close to the one achieved by the optimal solution (103 kbit/s) and better than the Greedy algorithm (90 kbit/s). In terms of user and group fairness, the three algorithms perform the same.


\section{Conclusion} \label{sec7}

In this paper, a LEO satellite-base NB-IoT system has been considered. To minimize the message exchange between the BS and the UEs, counteracting the increased RTD in the satellite channel, the amount of data transmitted by every user has been maximized determined by their specific channel conditions, data demands and NB-IoT standard limitations. Also, for the purpose of accommodating the high dynamicity of the system, a separation of access and data phases has been assumed, allowing for "short-term" resource allocation decisions with the latest and most accurate estimated parameters by the BS. A profit function has been designed which assigns different priorities to users depending only on the known parameters by the BS. The problem has been modeled as a 0-1 MKP which selects the set of users to be scheduled together in the data phase, maximizing the sum profit of the scheduled users, and minimizing the radio resource wastage. In order to tackle the differential Doppler problem, groups of users have been created and the available resources for data transmission have been optimally divided. Exploiting the unique features of our system, an Approximate algorithm for solving the 0-1 MKP has been proposed with a near-optimal performance and very low complexity. Finally, three KPIs have been designed with different performance targets and their optimal operational points have been found through numerical simulations. 

The overall framework proposed in this paper to aggregate the IoT traffic through a LEO satellite is novel. There are several directions of extension of this work. First of all, the differential Doppler problem that we tackle through the resource allocation strategy will also impact the access phase. Therefore, in the future work, we will jointly consider both transmission phases and propose a strategy which counteract the differential Doppler problem also in the access phase. Second, in this paper we assume that the access phase is designed in such a way that enables enough UEs to compete for the limited resource in the data phase. However, this may not always be the case since the IoT user distribution and density will change with the LEO satellite movement. Hence, in the future work we will address this aspect and create a model able to dynamically change the access phase duration and periodicity adapting to the IoT traffic. Last but not least, designing a unified system that includes both, the access and data phase, creates the possibility of working in a satellite constellation design targeting a global NB-IoT coverage.

\ifCLASSOPTIONcaptionsoff
  \newpage
\fi

\bibliographystyle{IEEEtran}
\bibliography{IEEEabrv,bibfile2}

\begin{thebibliography}{10}
\providecommand{\url}[1]{#1}
\csname url@samestyle\endcsname
\providecommand{\newblock}{\relax}
\providecommand{\bibinfo}[2]{#2}
\providecommand{\BIBentrySTDinterwordspacing}{\spaceskip=0pt\relax}
\providecommand{\BIBentryALTinterwordstretchfactor}{4}
\providecommand{\BIBentryALTinterwordspacing}{\spaceskip=\fontdimen2\font plus
\BIBentryALTinterwordstretchfactor\fontdimen3\font minus
  \fontdimen4\font\relax}
\providecommand{\BIBforeignlanguage}[2]{{%
\expandafter\ifx\csname l@#1\endcsname\relax
\typeout{** WARNING: IEEEtran.bst: No hyphenation pattern has been}%
\typeout{** loaded for the language `#1'. Using the pattern for}%
\typeout{** the default language instead.}%
\else
\language=\csname l@#1\endcsname
\fi
#2}}
\providecommand{\BIBdecl}{\relax}
\BIBdecl

\bibitem{ref10}
{3GPP TR 38.811}, ``{Technical Specification Group Radio Access Network; Study
  on New Radio (NR) to support non-terrestrial networks (Release 15) },'' Tech.
  Rep., September 2019.

\bibitem{ref1}
{3GPP TR 38.821}, ``{Technical Specification Group Radio Access Network;
  Solutions for NR to support non-terrestrial networks (NTN) (Release 16) },''
  Tech. Rep., December 2019.

\bibitem{ref11}
{3GPP RP-193235}, ``{Study on NB-Io/eMTC support for Non-Terrestrial Network
  },'' Tech. Rep., December 2019.

\bibitem{ref12}
S.~{Cioni}, R.~{De Gaudenzi}, O.~{Del Rio Herrero}, and N.~{Girault}, ``{On the
  Satellite Role in the Era of 5G Massive Machine Type Communications},''
  \emph{IEEE Network}, vol.~32, no.~5, pp. 54--61, Sep. 2018.

\bibitem{ref13}
\BIBentryALTinterwordspacing
N.~Alagha, ``{Satellite Air Interface Evolutions in the 5G and IoT Era},''
  \emph{SIGMETRICS Perform. Eval. Rev.}, vol.~46, no.~3, p. 93–95, Jan. 2019.
  [Online]. Available: \url{https://doi.org/10.1145/3308897.3308941}
\BIBentrySTDinterwordspacing

\bibitem{ref14}
A.~{Guidotti} \emph{et~al.}, ``{Architectures and Key Technical Challenges for
  5G Systems Incorporating Satellites},'' \emph{IEEE Transactions on Vehicular
  Technology}, vol.~68, no.~3, pp. 2624--2639, March 2019.

\bibitem{ref9}
O.~{Kodheli} \emph{et~al.}, ``{Resource Allocation Approach for Differential
  Doppler Reduction in NB-IoT over LEO Satellite},'' in \emph{2018 9th Advanced
  Satellite Multimedia Systems Conference and the 15th Signal Processing for
  Space Communications Workshop (ASMS/SPSC)}, Sep. 2018, pp. 1--8.

\bibitem{ref18}
------, ``{An Uplink UE Group-Based Scheduling Technique for 5G mMTC Systems
  Over LEO Satellite},'' \emph{IEEE Access}, vol.~7, pp. 67\,413--67\,427,
  2019.

\bibitem{ref32}
B.~Deng, C.~Jiang, H.~Yao, S.~Guo, and S.~Zhao, ``The next generation
  heterogeneous satellite communication networks: Integration of resource
  management and deep reinforcement learning,'' \emph{IEEE Wireless
  Communications}, vol.~27, no.~2, pp. 105--111, 2020.

\bibitem{ref33}
X.~Zhu, C.~Jiang, L.~Kuang, N.~Ge, and J.~Lu, ``Non-orthogonal multiple access
  based integrated terrestrial-satellite networks,'' \emph{IEEE Journal on
  Selected Areas in Communications}, vol.~35, no.~10, pp. 2253--2267, 2017.

\bibitem{ref19}
\BIBentryALTinterwordspacing
J.~Lee and J.~Lee, ``{Prediction-Based Energy Saving Mechanism in 3GPP NB-IoT
  Networks},'' \emph{Sensors}, vol.~17, no.~9, 2017. [Online]. Available:
  \url{https://www.mdpi.com/1424-8220/17/9/2008}
\BIBentrySTDinterwordspacing

\bibitem{ref20}
H.~{Malik} \emph{et~al.}, ``{Radio Resource Management Scheme in NB-IoT
  Systems},'' \emph{IEEE Access}, vol.~6, pp. 15\,051--15\,064, 2018.

\bibitem{ref21}
C.~{Yu} \emph{et~al.}, ``{Uplink Scheduling and Link Adaptation for Narrowband
  Internet of Things Systems},'' \emph{IEEE Access}, vol.~5, pp. 1724--1734,
  2017.

\bibitem{ref22}
Y.~{Yu} and J.~{Wang}, ``Uplink resource allocation for narrowband internet of
  things (nb-iot) cellular networks,'' in \emph{2018 Asia-Pacific Signal and
  Information Processing Association Annual Summit and Conference (APSIPA
  ASC)}, 2018, pp. 466--471.

\bibitem{ref23}
A.~{Azari} \emph{et~al.}, ``{On the Latency-Energy Performance of NB-IoT
  Systems in Providing Wide-Area IoT Connectivity},'' \emph{IEEE Transactions
  on Green Communications and Networking}, vol.~4, no.~1, pp. 57--68, 2020.

\bibitem{ref15}
J.~{Doré} and V.~{Berg}, ``{TURBO-FSK: A 5G NB-IoT EVOLUTION FOR LEO SATELLITE
  NETWORKS},'' in \emph{2018 IEEE Global Conference on Signal and Information
  Processing (GlobalSIP)}, Nov 2018, pp. 1040--1044.

\bibitem{ref17}
S.~{Cluzel} \emph{et~al.}, ``{3GPP NB-IOT Coverage Extension Using LEO
  Satellites},'' in \emph{2018 IEEE 87th Vehicular Technology Conference (VTC
  Spring)}, June 2018, pp. 1--5.

\bibitem{ref24}
L.~{Feltrin} \emph{et~al.}, ``{Narrowband IoT: A Survey on Downlink and Uplink
  Perspectives},'' \emph{IEEE Wireless Communications}, vol.~26, no.~1, pp.
  78--86, 2019.

\bibitem{ref5}
{3GPP TS 36.213}, ``{LTE; Evolved Universal Terrestrial Radio Access (E-UTRA);
  Physical layer procedures (version 15.7.0 Release 15) },'' Tech. Rep.,
  October 2019.

\bibitem{ref6}
A.~{Guidotti} \emph{et~al.}, ``{Satellite-enabled LTE systems in LEO
  constellations},'' in \emph{2017 IEEE International Conference on
  Communications Workshops (ICC Workshops)}, May 2017, pp. 876--881.

\bibitem{ref25}
O.~Kodheli \emph{et~al.}, ``Link budget analysis for satellite-based narrowband
  iot systems,'' in \emph{Ad-Hoc, Mobile, and Wireless Networks}.\hskip 1em
  plus 0.5em minus 0.4em\relax Cham: Springer International Publishing, 2019,
  pp. 259--271.

\bibitem{ref26}
\BIBentryALTinterwordspacing
J.~E. Beasley, ``An exact two-dimensional non-guillotine cutting tree search
  procedure,'' \emph{Operations Research}, vol.~33, no.~1, pp. 49--64, 1985.
  [Online]. Available: \url{http://www.jstor.org/stable/170866}
\BIBentrySTDinterwordspacing

\bibitem{ref27}
\BIBentryALTinterwordspacing
N.~Christofides and E.~Hadjiconstantinou, ``An exact algorithm for orthogonal
  2-d cutting problems using guillotine cuts,'' \emph{European Journal of
  Operational Research}, vol.~83, no.~1, pp. 21 -- 38, 1995. [Online].
  Available:
  \url{http://www.sciencedirect.com/science/article/pii/0377221793E02775}
\BIBentrySTDinterwordspacing

\bibitem{ref28}
\BIBentryALTinterwordspacing
S.~P. Fekete, J.~Schepers, and J.~C. van~der Veen, ``{An Exact Algorithm for
  Higher-Dimensional Orthogonal Packing},'' \emph{Operations Research},
  vol.~55, no.~3, pp. 569--587, 2007. [Online]. Available:
  \url{https://doi.org/10.1287/opre.1060.0369}
\BIBentrySTDinterwordspacing

\bibitem{ref29}
\BIBentryALTinterwordspacing
D.~Pisinger, ``An exact algorithm for large multiple knapsack problems,''
  \emph{European Journal of Operational Research}, vol. 114, no.~3, pp. 528 --
  541, 1999. [Online]. Available:
  \url{http://www.sciencedirect.com/science/article/pii/S0377221798001209}
\BIBentrySTDinterwordspacing

\bibitem{ref31}
S.~Martello and P.~Toth, \emph{Knapsack Problems: Algorithms and Computer
  Implementations}.\hskip 1em plus 0.5em minus 0.4em\relax USA: John Wiley \&
  Sons, Inc., 1990.

\bibitem{ref30}
{3GPP TR 45.820 }, ``{Cellular system support for ultra-low complexity and low
  throughput Internet of Things (CIoT) (V13.1.0) },'' Tech. Rep., November
  2015.

\end{thebibliography}


\end{document}